# Spontaneous down conversion of surface plasmon polaritons: strong-field consideration

VLADIMIR HIZHNYAKOV[(a)] and ARDI LOOT

*Institute of Physics, University of Tartu, W. Ostwaldi St. 1, 50411 Tartu, Estonia*

**Abstract** – A non-perturbative theory of the spontaneous down conversion (SDC) of surface plasmon polaritons at a metal-dielectric interface is presented. It is shown that the process is resonantly enhanced for the characteristic power of excitation, typically of the order of tens of watts. At a stronger excitation the yield of SDC decreases rapidly. At a stronger excitation the yield of SDC decreases rapidly. The reason for this decrease is the high rate of the change of surface plasmon polaritons by the laser field, exceeding the rate of the zero-point fluctuations responsible for the SDC process. The obtained results may help one to construct miniature sources of entangled photons for quantum communication.

## Introduction

Spontaneous down conversion (SDC) - the decay of a single photon into two photons of smaller frequencies [1,2] is a nonlinear quantum phenomenon which takes place due to the existence of zero-point electromagnetic fluctuations: laser light periodically perturbs the zero-point state of the medium and these perturbations cause a two-photon emission. In its origin SDC has similarity with the dynamical Casimir effect [3]: the two-photon emission arising from the change of the optical length of a resonator in the time which also causes the time-dependent perturbation of the zero-point state. SDC is of a remarkable interest from practical point of view, as the created photon pairs are in the entangled state and they may be used for quantum communication. At usual conditions SDC is characterized by the yield which depends only on the medium parameters [1,2]; therefore, this process is called the spontaneous parametric down conversion (SPDC). Ordinarily the yield of SPDC is of the order of $10^{-12}$ or less, i.e. the process has an extremely low probability. An exception is the down conversion at a metal-dielectric interface where, due to the field enhancement for the surface plasmon polaritons, one may expect to get a few orders of magnitude larger yield [4].

The independence of the yield of SDC from the excitation intensity should be observed as far as the perturbation theory for the description of the interaction of the laser light with the nonlinear medium is applicable. Usually in dielectrics this theory is violated if the excitation field $E$ becomes comparable or stronger than the atomic field $E_a \sim 10^{10}$ V/m. For such a strong excitation the optical breakdown can easily take place, which essentially complicates the study of the strong-field effect in the nonlinear optics. However, surface plasmon polaritons SPPs in metal-dielectric interfaces create a new situation: due to the field enhancement the characteristic limiting field for the perturbation theory may in this case be several orders of magnitude less than $E_a$. Therefore, the usual perturbation theory may not work for SPPs already for rather a moderate excitation, which is incomparably weaker than the excitations which may cause an optical breakdown. This means that one can readily meet the experimental conditions where the non-perturbative theory of SDC is required.

In this communication we are presenting such a non-perturbative theory. The approach used here is based on the theory of quantum vacuum effects in strong fields (see, e.g. Refs. [5,6]). The problem under consideration has a similarity with the two-phonon decay of a strong vibration in solids, which has been investigated in Ref. [7]. Using the method proposed in Ref. [7], we have found that there exists a critical power of excitation, typically of the order of 10 W, for which the yield of SDC is resonantly enhanced. If the power of excitation exceeds the critical one, then the yield of SDC diminishes. Such unexpected dependence of SDC on the excitation power has a clear physical explanation: in the case of strong excitation the laser field changes SPPs so rapidly that the zero-point fluctuations of SPPs, causing SDC, cannot follow these changes. This effect has a close analogy with the enhancement and the subsequent diminishing of the rate of two-phonon decay of a strong vibration in a crystal, with increasing its amplitude: at the characteristic amplitude typically of the order of few tenths of pm this rate gets a maximum; at larger amplitudes it becomes small [7-10]; see Ref [11], where this effect was verified experimentally. An

[(a)]E-mail: hizh@ut.ee

V. Hizhnyakov *et al.*

analogous enhancement of the two-quantum emission was also predicted [12,13] for the dynamical Casimir effect.

## Nonlinear wave equation for SDC

Here we are studying the SDC at the interface of a metal and a dielectric with nonzero second-order nonlinear susceptibility $\chi^{(2)}_{\alpha\beta\gamma}$. We are considering the case of the excitation of the interface by a quasi-monochromatic laser pulse. The scheme of the possible set-ups is given in Fig. 1. The created SPPs quanta are emitted to vacuum as photons through the Kretschmann prism. The SPPs in the interface are dominantly polarized perpendicularly to the interface. Therefore, to create the SPPs the dielectric should be oriented (cut) so that it has the non-zero components in this direction. E.g. in the case of a KDP crystal with nonzero $d_{14} = d_{25}$- and $d_{36}$-components of the tensor $\chi^{(2)}_{\alpha\beta\gamma}$ its symmetry plane should be oriented normally to the interface (here the Kleinman notations are used). Then the SDC will take place due to the $d_{14}$ component of $\chi^{(2)}$.

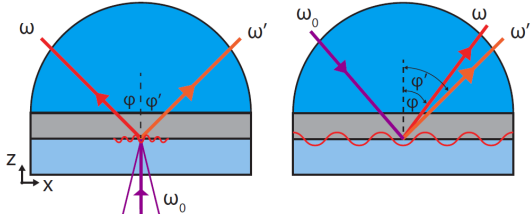

Fig. 1. SDC at metal-dielectric interface; left - subwavelength excitation through the SNOM tip; right- oblique excitation. Bottom layer is the dielectric with nonzero $\chi^{(2)}$, middle layer is metallic, the half-sphere with high refractive index allow to emit photons to vacuum; $\omega_0$, $\omega$ and $\omega' = \omega_0 - \omega$ are the frequencies of the excitation and of the emitted photons.

Down conversion can take place only if there already exists a nonzero field corresponding to at least one of the two generated quanta. In the case of SDC such a field is provided by zero-point fluctuations. To account these fluctuations, we need to consider the process quantum-mechanically. Therefore, to describe the SDC, we use the nonlinear wave-equation for operators

$$c^2 \nabla^2 \hat{E} - \partial^2 \hat{E}/\partial t^2 = n^{-2} \partial^2 \hat{P}^{NL}/\partial t^2 \qquad (1)$$

Here $\hat{E}$ is the operator of the electric field of one created quantum of the SPP, $\hat{P}^{NL}$ is the operator of the nonlinear polarization, $c = c_0/n$, $c_0$ is the velocity of light in vacuum, $n$ is the refractive index, the small divergence $(\nabla \hat{E})$ is neglected. We are considering the second-order nonlinear polarization operator

$$\hat{P}^{NL}(t) \cong 2\chi^{(2)} E_0(\vec{r},t) \hat{E}_1(\vec{r},t) \qquad (2)$$

where $E_0(\vec{r},t) = E_0(\vec{r})\cos(\omega_0 t)$ is the monochromatic laser field, $\omega_0$ is the frequency of this field, $\hat{E}_1(\vec{r},t)$ is the field operator of the second quantum of the SPP, which is also created in the SDC process, $\chi^{(2)} \equiv \chi^{(2)}_0 \eta_0 \eta \eta'$, $\chi^{(2)}_0$ is the working component of the second-order nonlinear susceptibility of the dielectric, $\eta_0$ is the renormalization factor of the laser field $E_0$ at the interface which creates the moving polarization along the interface, $\eta$ and $\eta'$ are the field enhancement factors for the created SPP quanta. The physical meaning of the right-hand side term in Eq. (1) is the quantum force acting to SPPs which arises from the common action of the laser field and zero-point fluctuations of SPPs.

The typical value of $\chi^{(2)}_0$ for dielectrics is $\sim 1$ pm/V. The field enhancement factor for $\lambda = 1100$ nm in case of 60 nm silver film is $\eta \approx 20$ (see Fig. 2). Note that for long-range propagating SPPs the enhancement factor may be several times larger [14]. The renormalization of the laser field which creates the moving polarization along the interface is usually not large, i.e. $\eta_0 \sim 1$ [15]. Therefore, one can estimate for usual SPPs in the interface $\chi^{(2)} \sim 0.4$ nm/V.

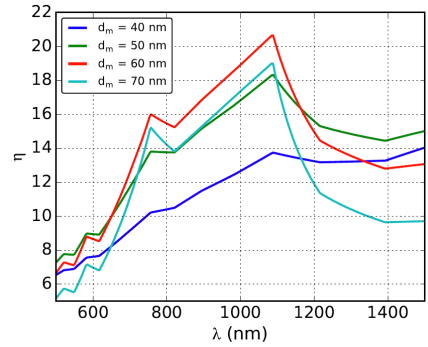

Fig. 2. The enhancement of the field of a SPP in structure consisting of prism (refractive index 2.0) silver film with thickness $d_m$ and dielectric (refractive index 1.5) for wavelength $\lambda$ (the refractive index of silver from [16]).

## Equation for mode operator

Let us expand $\hat{E}$ into the series of plane wave operators:

$$\hat{E}(\hat{r},t) = \sum_{\vec{k}} \hat{E}_{\vec{k}}(t) e^{i\vec{k}\vec{r}} \qquad (3)$$

Here $\hat{E}_{\vec{k}}$ is the mode operator, $\hat{a}_{\vec{k}}$ and $\hat{a}^+_{\vec{k}}$ are the destruction and creation operators, $S$ is the area of the metal-dielectric interface, $\omega_k = ck$ is the frequency of the SPP with the wave vector $\vec{k}$. Substituting this equation into Eq. (1) we get the following equation for the mode operator:

$$\ddot{\hat{E}}_{\vec{k}}(t)+\omega_k^2\hat{E}_{\vec{k}}(t)=\left(4\omega_k^2\upsilon s_0/S\right)\sum_{\vec{k}'}\Psi_{\vec{k},\vec{k}'}\hat{E}_{\vec{k}'}(t) \quad (4)$$

where $\upsilon=\chi^{(2)}E_0/2n^2$ is the dimensionless interaction parameter,

$$\Psi_{\vec{k},\vec{k}'}=(s_0 E_0)^{-1}\int dS e^{-i(\vec{k}+\vec{k}')\vec{r}}E_0(\vec{r}) \quad (5)$$

$s_0=\int dS E_0(\vec{r})$ (the integration is performed along the interface). Here we take into account that for the mode $\vec{k}$ the second time derivative in the right-hand side of Eq. (1) can be replaced by $-\omega_k^2$. Using the Green's function of the harmonic oscillator $\omega^{-1}\Theta(t)\sin(\omega t)$, we can present Eq. (4) in the integral form

$$\hat{E}_{\vec{k}}(t)\approx\hat{E}_{\vec{k}}^{(0)}(t)+\left(4\upsilon\omega_k s_0/S\right)\times$$
$$\times\int_{-\infty}^{t}\sin\left(\omega_k(t-t')\right)\cos(\omega_0 t')\sum_{\vec{k}'}\Psi_{\vec{k},\vec{k}'}\hat{E}_{\vec{k}'}(t')dt' \quad (6)$$

where $\hat{E}_{\vec{k}}^{(0)}(t)=i\sqrt{\hbar\omega_k/2\varepsilon_0}\left(\hat{a}_{\vec{k}}^{(0)}e^{-i\omega_k t}-\hat{a}_{-\vec{k}}^{(0)+}e^{i\omega_k t}\right)$ is the initial operator of the mode $\vec{k}$, $\hat{a}_{\vec{k}}^{(0)}$ is the initial destruction operator of this mode, $\vec{k}'$ corresponds to the momentum of the second emitted SPP quantum.

We suppose that the excitation is done by a Gaussian-like beam giving

$$E_0(\vec{r})=E_0 e^{i\vec{k}_0\vec{r}-r^2/2r_0^2} \quad (7)$$

where $r_0$ is the size of the excited area, $\vec{k}_0$ accounts for an oblique excitation. In this case $s_0=2\pi r_0^2$ and

$$\Psi_{\vec{k},\vec{k}'}=e^{-r_0^2|\vec{k}_0-\vec{k}-\vec{k}'|^2/2} \quad (8)$$

We consider that $r_0$ is smaller than the propagation length of SPPs $l_p$. For 60 nm silver film in visible (800 nm) the propagation length is of the order of 0.3 mm, however for long-range propagating SPPs $l_p$ may exceed 1 cm [16].

## Number of the emitted SPP quanta

In the Heisenberg picture the spontaneous emission of SPP quanta is determined by the transformation of their operators in time. To find the number of the emitted quanta of SPPs one may apply the Coleman theorem [17] (see also [5,6]) according to which a classical time-dependent field leads to the Bogoliubov transformation of the creation and destruction operators. In the $t\to\infty$ limit the transformed destruction operator gets the form

$$\hat{a}_{\vec{k}}(t)=\mu_{\vec{k}}\hat{a}_{\vec{k}}^{(0)}e^{-i\omega_k t}+\nu_{\vec{k}}\hat{a}_{-\vec{k}}^{(0)+}e^{i\omega_k t} \quad (9)$$

$\mu_{\vec{k}}$ and $\nu_{\vec{k}}$ are the parameters of the transformation, the created operator is transformed analogously. Therefore the vacuum states for initial operators $|0\rangle$ is not that for the final operators: the final particles are present in this state [5,6]. The number of particles equals to $N_{\vec{k}}=\langle 0|\hat{a}_{\vec{k}}^+\hat{a}_{\vec{k}}|0\rangle=|\nu_{\vec{k}}|^2$, From these relations it follows that the rate $\dot{N}_{\vec{k}}\equiv dN_{\vec{k}}/dt$ can be obtained from the following large-time asymptotic of the field correlation function [10]:

$$N_{\vec{k}}=\frac{2\varepsilon_0}{\hbar\omega_k}e^{-i\omega_k\tau}\langle 0|\hat{E}_{\vec{k}}(t+\tau)\hat{E}_{\vec{k}}(t)|0\rangle,\; t\to\infty \quad (10)$$

This asymptotical value can be found if Eq. (6) is solved.

## Weak excitation

If the laser excitation is weak then in Eq. (6) one can replace the operator $\hat{E}_{\vec{k}'}(t')$ under the integral by its non-perturbed value $\hat{E}_{\vec{k}'}^{(0)}(t')$. Taking into account that in the $t\to\infty$ limit the factor $\sin(\omega_k(t-t'))\cos(\omega_0 t')$ can be replaced by $e^{i(\omega_k t+(\omega_0-\omega_k)t')}/4$ one gets

$$\hat{E}_{\vec{k}}(t)\simeq\hat{E}_{\vec{k}}^{(0)}(t)+\frac{\upsilon\omega_k s_0}{S}\int_{-\infty}^{t}e^{i\omega_k(t-t')+i\omega_0 t'}\sum_{\vec{k}'}\Psi_{\vec{k},\vec{k}'}\hat{E}_{\vec{k}'}^{(0)}(t')dt'$$

Let us substitute this equation into Eq. (10) and take the time derivative. We get

$$\dot{N}_{\vec{k}}^{(0)}=2\pi\omega_k\frac{|\upsilon|^2 s_0^2}{S^2}\sum_{\vec{k}'}\omega_{k'}|\Psi_{\vec{k},\vec{k}'}|^2\delta(\omega_0-\omega_k-\omega_{k'}) \quad (11)$$

The sum in Eq. (11) should be replaced by integration over the frequency $\omega$ and angle $\varphi$:

$$\sum_{\vec{k}}\to\sum_{k,\varphi}\to\left(S/2\pi^2 c^2\right)\int_0^{\infty}\omega d\omega\int_0^{2\pi}d\varphi \quad (12)$$

Then we get the following equation for the rate of emission of SPP quanta with the frequencies $\omega$ and $\omega'=\omega_0-\omega$ in the directions $\varphi$ and $-\varphi'$:

$$\dot{N}^{(0)}(\omega,\varphi,\varphi')=\frac{|\nu|^2}{2\pi\omega_0^4}\omega^2(\omega_0-\omega)^2\Psi^2(\omega,\varphi,\varphi') \quad (13)$$

where

$$\nu=4\pi^2 r_0^2 E_0\chi^{(2)}/\lambda_0^2 n^2 \quad (14)$$

is the dimensionless energy of the nonlinear interaction of the SPP quantum with the laser light,

$$\Psi^2(\omega,\varphi,\varphi')=\exp[(r_0^2/c^2)((\omega\sin\varphi-\omega'\sin\varphi')^2+ \\ +(k_0 c-\omega\cos\varphi-\omega'\cos\varphi')^2] \quad (15)$$

describes the angular dependence of the emission.

## Strong excitation

In the case of the strong excitation the approximation $\hat{E}_{\vec{k}'}(t')\approx\hat{E}_{\vec{k}'}^{(0)}(t')$ cannot be used in the right-hand side of Eq.(6). In this case, in general, Eq. (6) cannot be solved. However, if $\Psi_{\vec{k},\vec{k}'}$ is a constant or if it can be factorized then the solution may be found. Below we consider two such cases.

**a. Subwavelength excitation.** This excitation can be realized by using a scanning near-field optical microscopy

V. Hizhnyakov *et al.*

(SNOM) tip (see Fig. 1, left). In this case $\Psi \simeq 1$ and the emission does not depend on the direction. Using the totally-symmetric mode operators $\hat{E}_k = \left(1/\sqrt{2\pi}\right)\sum_\varphi \hat{E}_{k,\varphi}$ we get from Eq.(6) the relation

$$\hat{E}_k(t) \cong \hat{E}_k^{(0)}(t) + \left(16\pi^2 \omega_k r_0^2 \upsilon / \sqrt{S}\right) \times \\ \times \int_{-\infty}^{t} \sin(\omega_k(t-t'))\cos(\omega_0 t')\hat{Q}(t')dt' \quad (16)$$

where $\hat{Q}(t') = S^{-1/2}\sum_{k'}\hat{E}_{k'}(t')$. Taking into account that in the $t \to \infty$ limit only the term $e^{i\omega_k(t-t')+i\omega_0 t'}/4$ of $\sin(\omega_k(t-t'))\cos(\omega_0 t')$ gives a contribution we get

$$\dot{N}_k \simeq \left(32\pi^4 r_0^4 \omega_k |\upsilon|^2/\hbar S\right)\int_{-\infty}^{\infty} dt_1 e^{i(\omega_0-\omega_k)t_1} D(t_1) \quad (17)$$

where

$$D(t_1) = D(t+t_1, t) = \langle 0|\hat{Q}(t+t_1)\hat{Q}(t)|0\rangle, \ t \to \infty \quad (18)$$

depends on the time difference. Taking into account Eq. (12), we get the spectral rate of the emission in the form

$$J(\omega) \equiv \dot{N}(\omega) = \left(32\pi^3 |\upsilon|^2 \varepsilon_0 r_0^4 \omega^2/c^2\hbar\right)D(\omega_0 - \omega) \quad (19)$$

where $D(\omega) = \int_{-\infty}^{\infty}dt e^{i\omega t}D(t)$. For small $\upsilon$ one gets

$$D(\omega) \approx D^{(0)}(\omega) = \hbar\omega^2/4\pi^2 c^2 \varepsilon_0 \quad (20)$$

Note that only positive frequencies of $D(t)$ give a contribution to $N_k$. This allows one to consider only the positive frequency part of $\hat{Q}(t)$.

To find $D(\omega)$ for an arbitrary $\upsilon$ we proceed from Eq.(16). Taking into account that only the item $e^{i\omega_0 t'}/2$ gives a contribution to the positive frequency operator, we get the following equation for $\hat{Q}(t)$:

$$\hat{Q}(t) \cong \hat{Q}^{(0)}(t) + \upsilon\int_{-\infty}^{\infty}dt' G(t-t')e^{i\omega_0 t'}\hat{Q}(t') \quad (21)$$

Here $G(t) = \Theta(t)\int_0^{\omega_0}(\omega/\omega_0)^2 \sin(\omega t)d\omega$, $\Theta(t)$ is the Heaviside step function. For further consideration it is convenient to present Eq. (21) in the form

$$\hat{Q}(t) = \hat{Q}^{(0)}(t) + \upsilon\int_{-\infty}^{\infty}dt' G(t-t')e^{i\omega_0 t'}\hat{Q}^{(0)}(t') + \\ + \upsilon^2 \int_{-\infty}^{\infty}dt' \int_{-\infty}^{\infty}dt_1' G(t-t')G(t'-t_1')e^{i\omega_0(t'-t_1')}\hat{Q}(t_1') \quad (22)$$

Here the $\propto \upsilon$ term, due to the factor $e^{i\omega_0 t'}$, oscillates fast and in the larger $t, t'$ limit averages out. The $\propto \upsilon^2$ term is essential, as the term $e^{i\omega_0(t_1-t_1')}$ for large $t', t_1' \sim t$ but finite $t' - t_1'$ does not average out. Substituting now Eq. (22) into Eq. (18) and neglecting the $\propto \upsilon$ term we get the following equation:

$$D(t-t') \simeq D_0(t-t') + \\ + \upsilon^2 \int_{-\infty}^{\infty}dt_1 \int_{-\infty}^{\infty}dt_1' G(t-t_1)G(t_1-t_1')e^{i\omega_0(t_1-t_1')}D(t_1'-t') \quad (23)$$

where $D_0(t-t') = \langle 0|\hat{Q}^{(0)}(t)\hat{Q}(t')|0\rangle$. The latter function satisfies the same equation as $D(t'-t)$, but with $D^{(0)}(t-t') = \langle 0|\hat{Q}^{(0)}(t)\hat{Q}^{(0)}(t')|0\rangle$ instead of $D_0(t'-t)$. The Fourier transformation of Eq. (23) gives the linear equation

$$D(\omega) = D_0(\omega) + \upsilon^2 G(\omega)G^*(\omega_0-\omega)D(\omega) \quad (24)$$

where

$$G(\omega) = 1/2 + (\omega^2/2\omega_0^2)\left(\ln(\omega_0^2/\omega^2 - 1) + i\pi\right) \quad (25)$$

An analogous equation also exists for $D_0^*(\omega)$. Solution of these equations reads $D(\omega) = D^{(0)}(\omega)R(\omega)$, where

$$R(\omega) = \left|1 - |\upsilon|^2 G(\omega)G^*(\omega_0-\omega)\right|^{-2}$$

Consequently, the non-perturbative consideration of $D(\omega)$ results in the appearance of the additional resolvent factor $R(\omega)$. As a result one gets

$$J(\omega) \simeq \frac{2\pi|\upsilon|^2 \omega^2(\omega_0-\omega)^2/\omega_0^4}{\left|1 - |\upsilon|^2 G(\omega)G^*(\omega_0-\omega)\right|^2} \quad (26)$$

From this equation it follows that for small $\upsilon$ (small intensity of excitation $I_0$), as it should be, the emission rate linearly increases with $I_0 \propto |E_0|^2$. However for

$$\upsilon \approx \upsilon_r = |G(\omega_0/2)|^{-1} = 1.336 \quad (27)$$

the emission intensity gets for $\omega = \omega_0/2$ a maximum of a Lorentzian shape:

$$J(\omega) \simeq \frac{9/32\pi\upsilon_r^2}{(\omega/\omega_0 - 1/2)^2 + \alpha^2(\upsilon - \upsilon_r)^2} \quad (28)$$

where $\alpha = 3/4\pi\upsilon_r^3 \approx 0.10$. The value $\upsilon_r$ corresponds to the case when the energy of the nonlinear interaction of the SPP quantum with the laser light becomes comparable with its own energy. The total rate of the emission $I = \int \dot{N}(\omega)d\omega$ diverges as $|\upsilon - \upsilon_r|^{-1}$. For $\upsilon > \upsilon_r$ the rate turns to decrease. Such unexpected dependence of the SDC on the excitation intensity has a clear physical explanation: in this case the laser field perturbs SPPs so rapidly that the zero-point fluctuations of SPPs cannot follow the change of the field and the generation of the SPP quanta caused by these fluctuations cannot take place.

The obtained dependence of the rate on $\upsilon$ is analogous to the dependence of the two-phonon decay rate of a strong local vibration in a crystal on its amplitude found in Ref. [7]: this decay has maximal rate at a characteristic amplitude typically of the order of a few tenths of pm [8,9]; at larger amplitudes the rate slows-down. This-type of the dependence of the decay of the local vibration on its amplitude was experimentally observed in Ref. [11] for the relaxation of UV-light excited $Xe_2^*$ and $Kr_2^*$ dimers in solid $Xe$ and $Kr$, respectively, by means of hot luminescence.

Taking $\chi_0^{(2)} \sim 1$ pm/V (usual value for dielectrics), $r_0 \sim \lambda_0$ - the excited area close to the Abbe diffraction limit, $\eta_0 \sim 1$, $\eta = \eta_1 \approx 20$ (the field enhancement in the case of a silver film [13]), we find that $v = v_r$ is achieved for the excitation field $E_0 \sim 10^6$ V/cm and for the power $\sim 10$ W. For large $|v| > v_r$ the rate of the emission turns to decrease with the increase of the excitation intensity: for $|v| \gg v_r$ the rate of emission decreases as $I_0^{-1}$.

**b. Oblique incidence, significant excited area.** We are considering here the excitation of a metal-dielectric interface by a Gaussian-like beam with $k_0 = 2k_{\omega_0/2}$ and $r_0 \gg \lambda_0$. In this case (experimental scheme given on Fig. 1, right) the SDC corresponds to a process with phase-matching of laser-induced polarization with two created SPP quanta moving forward. In this case for $\omega \to \omega_0/2$ the factor $\Psi_{\vec{k},\vec{k}'}$ tends to unity like in the case of the subwavelength excitation. Therefore, also in this case, one can expect getting an enhancement of SDC for $\omega = \omega_0/2$.

In this case for $|\omega - \omega_0/2| < \Lambda_0^{-1/2}$ the factor $\Psi$ practically does not depend on $\omega$ (see Fig. 3) and equals

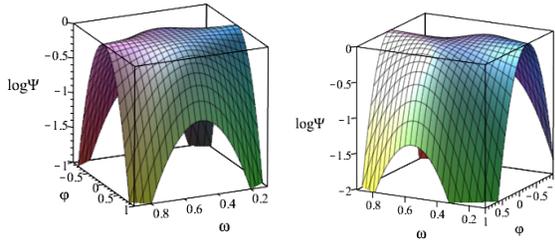

Fig. 3. Factor $\Psi$ for $\Lambda_0 = 10$, $\varphi' = \varphi$ (left) and $\varphi' = \varphi + 0.2$ (right).

$$\Psi(\varphi,\varphi') \approx e^{-\Lambda_0^2\left((\varphi-\varphi')^2 + \varphi^2\varphi'^2\right)} \quad (29)$$

where $\Lambda_0 = \pi r_0/\lambda_0 \gg 1$. Then Eq. (6) gets the form

$$\hat{E}_{k,\varphi}(t) \approx \hat{E}_{k,\varphi}^{(0)}(t) + (16\pi^2\omega_k r_0^2 v/S) \times$$
$$\times \int_{-\infty}^{t} \sin(\omega_k(t-t'))\cos(\omega_0 t)\sum_{\varphi'}\Psi(\varphi,\varphi')\hat{Q}_{\varphi'}(t') \quad (30)$$

where $\hat{Q}_{\varphi'}(t') = \sum_{k'}\hat{E}_{k',\varphi'}$ satisfies the equation

$$\hat{Q}_\varphi(t) \cong \hat{Q}_\varphi^{(0)}(t) +$$
$$+ v\int_{-\infty}^{\infty} dt'G(t-t')e^{i\omega_0 t'}\sum_{\varphi'}\Psi(\varphi,\varphi')\hat{Q}_{\varphi'}(t') \quad (31)$$

These equations are analogous to Eqs. (16) and (21) but they include the dependence on $\varphi$ and the additional sum over $\varphi'$. The rate of emission in $\varphi$ direction is also analogous to Eq. (19):

$$\dot{N}_\varphi(\omega) \cong \frac{4\pi\varepsilon_0}{c^2\hbar}(vs_0\omega)^2\sum_{\varphi'}\Psi^2(\varphi,\varphi')D_{\varphi'}(\omega_0-\omega) \quad (32)$$

where $D_{\varphi'}(\omega)$ is the Fourier transform of the correlation function $D_{\varphi'}(t,t') = \langle 0|\hat{Q}_{\varphi'}(t)\hat{Q}_{\varphi'}(t')|0\rangle$. The function $\Psi^2(\varphi,\varphi')$ has a very sharp peak for $\varphi' = \varphi$ due to the Gaussian factor $e^{-2\Lambda_0^2(\varphi-\varphi')^2}$ (for this factor essential are the values $|\varphi-\varphi'| \lesssim \Lambda_0^{-1}$). This allows one to replace the factor $e^{-2\Lambda_0^2\varphi^2\varphi'^2}$ by $e^{-2\Lambda_0^2\varphi^4}$. Due to the same reason the function $D_{\varphi'}(\omega_0-\omega)$ in Eq. (32) can be replaced by $D_\varphi(\omega_0-\omega)$. This gives

$$\dot{N}_\varphi(\omega) \simeq \frac{\sqrt{2\pi^3}\varepsilon_0 v^2 s_0^2\omega^2}{\Lambda_0 c^2\hbar}e^{-2\Lambda_0\varphi^4}D_\varphi(\omega_0-\omega) \quad (33)$$

where $D_\varphi(\omega_0-\omega)$ is the Fourier transform of the function $D_\varphi(t-t')$, which satisfies analogous to Eq. (23) equation

$$D_\varphi(t-t') \simeq D_{0\varphi}(t-t') + v^2\sum_{\varphi',\varphi''}\Psi(\varphi,\varphi')\Psi(\varphi',\varphi'')\times$$
$$\times \int_{-\infty}^{\infty}dt_1\int_{\infty}^{\infty}dt_1'G(t-t_1)G(t_1-t_1')e^{i\omega_0(t_1-t_1')}D_{\varphi''}(t_1'-t')$$

Here we take once more into account the sharp peaks of the Gaussian factors $e^{-\Lambda_0^2(\varphi-\varphi')^2}$ and $e^{-\Lambda_0^2(\varphi'-\varphi'')^2}$ for $\varphi' = \varphi$ and $\varphi'' = \varphi'$, respectively. These sharp peaks allow one to replace the factor $e^{-\Lambda_0^2\varphi'^2(\varphi^2+\varphi''^2)}$ in $\Psi(\varphi,\varphi')\Psi(\varphi',\varphi'')$ by $e^{-2\Lambda_0^2\varphi^4}$ and the function $D_{\varphi''}(t_1'-t')$ by $D_\varphi(t_1'-t')$. Taking then into account the relation

$$\sum_{\varphi',\varphi''}e^{-\Lambda_0^2\left((\varphi-\varphi'')^2+(\varphi''-\varphi')^2\right)} \cong \pi/\Lambda_0^2$$

we get the equation

$$D_\varphi(t-t') \simeq D_{0\varphi}(t-t') +$$
$$+ v_\varphi^2\int_{-\infty}^{\infty}dt_1\int_{\infty}^{\infty}dt_1'G(t-t_1)G(t_1-t_1')e^{i\omega_0(t_1-t_1')}D_\varphi(t_1'-t') \quad (34)$$

where $v_\varphi = \pi v\Lambda_0^{-1}e^{-\Lambda_0^2\varphi^4}$. Eq.(34) is analogous to Eq.(23). Therefore, the above-derived Eq.(28) applies also in this case, however for the emission in the direction $\varphi$. As a result, we get the spectral-angular rate of the emission of SPP quanta with the frequency $\omega$ in the direction $\varphi$ in the form

V. Hizhnyakov *et al.*

$$\dot{N}_\varphi(\omega) \simeq \frac{2\pi |v_\varphi|^2 \omega^2 (\omega_0 - \omega)^2 / \omega_0^4}{\left|1 - |v_\varphi|^2 G(\omega) G^*(\omega_0 - \omega)\right|^2} \quad (35)$$

The total spectral rate of the emission in this case is given by the integral $J(\omega) = \int d\varphi \dot{N}_\varphi(\omega)$. The dependence of $J$ on $\omega$ is presented in Fig. 4. For $\omega = \omega_0/2$ and for the excitation intensity $I_r$ corresponding to $v_{\varphi=0} = v_r = 1.336$ this rate is resonantly enhanced. The corresponding power

$$W_r = \pi I_r r_0^2 = \frac{v_r^2 n^4 \lambda_0^2}{64\pi^3 Z_0 \eta_0^2 \eta^4 |\chi^{(2)}|^2} \quad (36)$$

does not depend on $r_0$ (supposing that $l_p > r_0 > \lambda_0$).

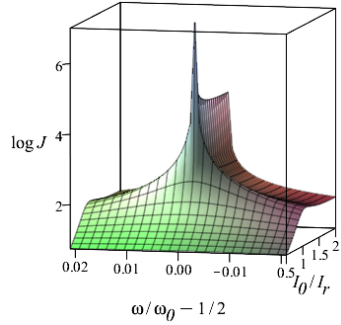

Fig. 4: Spectral rate $J$ (au) of SDC in the case of oblique excitation (see Fig. 1, right) with $k_0 = 2k_{\omega_0/2}$; $\omega$ is the emission frequency, $\omega_0$ and $I_0$ are the frequency and the intensity of excitation, $I_r = W_r/\pi r_0^2$.

The total rate of the emission $\Gamma = \int I(\omega) d\omega$ has also for $I_0 = I_r$ sharp peak (Fig. 5). For power $I_0 > I_r$ the rate of the emission under consideration turns to decrease. For $I_0 \gg I_r$ it decreases as $I_0^{-1}$, similarly to previously considered case of subwavelength excitation. The physical reason for such unexpected behaviour of the emission under consideration is also the same: in the case of strong excitation the laser field perturbs SPPs so rapidly that the zero-point fluctuations of SPPs cannot follow the change of the field and the generation of the SPP quanta cannot take place.

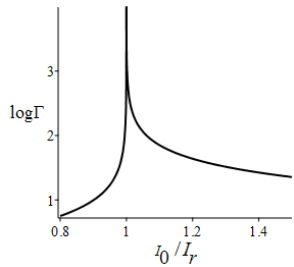

Fig. 5: Dependence of total rate $\Gamma$ (au) of SDC in case of oblique excitation with $k_0 = 2k_{\omega_0/2}$ on excitation intensity.

## Discussion

According to the presented theory two mechanisms can lead to the enhancement of SDC at the metal-dielectric interface: the amplification of the local electric field associated with SPPs and the resonant enhancement of the SDC process when the energy of the interaction of a SPP quantum with the laser light at the interface becomes comparable to the quantum energy.

Let us first estimate the enhancement of the yield $\zeta^{(0)}$ of the SDC in the weak excitation limit (in this case only the first mechanism works). The rate of the initial photons is equal to $\dot{N}_0 = \pi r_0^2 I_0/\hbar\omega_0$, where $I_0 = |E_0|^2/Z_0$ is the excitation intensity, $Z_0 = 376.7$ Ohm is the impedance of the free space. In the case of oblique excitation with the fulfilled phase-matching condition $k_0 = 2k_{\omega_0/2}$ the yield of SDC in the weak excitation limit equals to

$$\zeta^{(0)} = \frac{64\alpha\pi^2 \hbar \omega_0^2 \eta_0^2 \eta^4}{15n^4 \lambda_0^2} Z_0 |\chi^{(2)}|^2 \sqrt{\lambda_0/r_0} \quad (37)$$

Taking $r_0 \sim 10^3 \lambda_0$, $n = 1.5$, $\lambda_0 = 0.5\mu m$ we get the yield in the order of $\sim 4 \cdot 10^{-14} \eta^4$. Taking the field enhancement factor for SPPs $\eta \approx 20$ and $\eta_0 \sim 1$, one gets the yield $\zeta^{(0)} \sim 6 \cdot 10^{-9}$, which is several orders of magnitude larger than the yield in usual cases.

This yield may be additionally enhanced due to the second mechanism if the excitation power gets close to the resonant power $W_r$; for given above parameters $W_r \sim 20$ W. However, for the stronger excitation the yield will be reduced. We have shown that this reduction has a quantum origin: in the case of strong excitation the laser field perturbs SPPs so rapidly that the zero-point fluctuations of SPPs cannot follow the change of the field and, therefore, no generation of the SPP quanta caused by these fluctuations can take place.

One can additionally enlarge the yield of SDC corresponding to $W_r$ and reduce the critical power $W_r$ if to use a properly nanostructured interface, in the same way as it was done in Ref. [18] for four-wave mixing. Indeed, a grating of the interface with a period $a$ results in the appearance of SPPs with the same frequency but with a new wave numbers $k_n = k - nk_a$, $n = \pm 1, \pm 2, ...$ For a properly-chosen $a$ the phase matching condition will be fulfilled for the initially excited SPP with the frequency $\omega_0$ and for the SPPs with the frequency $\omega_0/2$ created in the SDC process. This will allow one to achieve a strong amplification $\eta_0 \gg 1$ of the field $E_0$ of the moving polarization and to amplify the working nonlinear interaction. This may allow one to additionally increase the yield of the SPDC more than two orders of magnitude.

## Conclusion

We have shown that SDC may be essentially enhanced if to use a properly prepared metal dielectric interface. There are two mechanisms which can lead to the enhancement: 1) amplification of the local electric field in the interface, resulting in an enlarged nonlinear interaction of the laser light with surface plasmon polaritons (we discussed this mechanism in Refs. [4,13]), 2) resonant enhancement of the SDC process when the change of the energy of the SPP quantum due to its nonlinear interaction with the laser light in the interface becomes comparable with its own energy. For the usual parameters of the interface this enhancement should take place for the excitation power in the order of 10 W. By using both of these, the enhancement allows one to get the yield of SDC many orders of magnitude larger than in the case of SDC in usual dielectrics. As a result, it may be possible to construct miniature devices serving as efficient sources of entangled photons, which may be used in quantum communication.

One can also expect that the non-perturbative theory of SDC presented here may be extended also for the non-perturbative description of other nonlinear optical processes, e.g. such as the four-wave mixing, the surface enhanced Raman scattering and the two-photon superradiance.

*\*\*\**

The research was supported by the Estonian research project IUT2-27.